\documentclass[runningheads]{llncs}
\usepackage[T1]{fontenc}
\usepackage{graphicx}
\usepackage{amsmath}
\usepackage{amssymb}
\usepackage{booktabs}
\usepackage{array}
\usepackage{tabularx}
\newcommand{\Etgt}{\mathbf{e}_{\mathrm{tgt}}}
\newcommand{\zI}{\mathbf{z}_{I}}
\newcommand{\cplx}[1]{\mathbb{C}^{#1}}
\newcommand{\real}[1]{\mathbb{R}^{#1}}
\begin{document}
\title{MuseFlow-TSE: Music-Aware Target Speech Extraction via Cross-Source Band-Splitting and Data-to-Data Flow Matching}
\titlerunning{MuseFlow-TSE}
\author{Lu Ma\thanks{Corresponding author.} \and Yongliang Wang \and Zhewen Su \and Xin Tian}
\authorrunning{L. Ma et al.}

\institute{JD.com, Inc., Beijing, China\\
\email{iamroad@163.com}}
\maketitle
\begin{abstract}
Target speech extraction (TSE) isolates a speaker's voice given an enrollment utterance, but existing work is validated almost exclusively without music. Discriminative methods are content-faithful yet masking-bounded; generative methods improve naturalness but may hallucinate, drift in identity, and require many steps. We present MuseFlow-TSE, a ``discriminative deletion, generative insertion'' framework for music-bearing scenes. Stage-1 is a music-aware cross-source band-split extractor that estimates target, interfering speakers, music, and noise under conservation and music-leakage constraints. Stage-2 is a data-to-data flow refiner initialized by the discriminative estimate, negatively conditioned on interference stems, and run for 1--4 difficulty-adaptive Euler steps with an end-to-end fidelity objective.
\end{abstract}
\section{Introduction}
Target speech extraction (TSE) isolates a desired speaker's voice from a mixture given an enrollment utterance. Real scenes, however, also contain background music, singing, noise, and reverberation, whereas existing TSE work is validated almost exclusively on music-free mixtures~\cite{mask2flow,solospeech,adflowtse,ddtse}, with clear degradation in zero-shot music evaluations~\cite{solospeech}. Music is wideband, strongly harmonic, and overlaps the voice band far more than noise---a distinct and difficult interferer.

Discriminative systems \emph{delete} interference but cannot restore target content; generative systems \emph{insert} detail but, from Gaussian noise, are slow and may hallucinate or drift in identity. D/I evidence shows that a masked start reduces the NFE from 16 to 1~\cite{mask2flow}, and discriminative estimates are effective refinement starts~\cite{ddtse,solospeech}. Music adds two challenges: wideband overlap increases insertion demand, requiring difficulty-adaptive steps, and deleted accompaniment or singing may be hallucinated back, causing identity drift~\cite{meanflowtse}.

We propose \textbf{MuseFlow-TSE} around \emph{discriminative deletion, generative insertion, explicit music modeling, and negative interference conditioning}. Stage-1 fuses BS-RoFormer band-splitting~\cite{bsroformer} with SepReformer-style early split and cross-source interaction~\cite{sepreformer}, estimating target/interferer/music/noise slots under conservation, grouped PIT, and a MusicLeak loss. Stage-2 starts from the discriminative estimate, uses the interference stems as negative conditions, adapts its step count with a difficulty head, and is trained end to end. A frozen ReDimNet encoder~\cite{redimnet} conditions only the target pathway. We describe the architecture, training, and nine ablations.

\section{Related Work}
Discriminative TSE uses speaker-conditioned masking~\cite{voicefilter,speakerbeam,spex}, embedding-free cues~\cite{useftse}, or separation backbones~\cite{tfgridnet,sepreformer}; these methods are subtractive and do not model music. Generative TSE includes diffusion and flow approaches~\cite{dpmtse,ddtse,solospeech,adflowtse,meanflowtse,mask2flow}, while SoloAudio's conditioned DiT informs Stage-2~\cite{soloaudio}. These systems are evaluated on music-free benchmarks and do not use discriminative interference estimates to constrain generation. Stage-1 adapts BS-RoFormer band-splitting~\cite{bsroformer} to speaker-conditioned extraction; Stage-2 builds on flow matching~\cite{lipman} and MeanFlow/$\alpha$-Flow~\cite{meanflow,alphaflow}.

\section{Method}
\subsection{Problem Formulation and Overall Framework}
Let the 16\,kHz single-channel mixture be $x = s + \sum_{j=1}^{J} u_j + m + n$, where $s$ is the target speech, $u_j$ are up to three interfering speakers, $m$ is music or singing, and $n$ is noise; reverberation is optional. Given an enrollment $e$, TSE estimates $\hat{s}\approx s$. We denote the complex STFTs of $x$ and $s$ by $X,S\in\cplx{F\times T}$, respectively, where $F$ is the number of retained frequency bins and $T$ is the number of time frames. The actual front-end uses a 512-sample Hann window, a 512-point FFT, and a 128-sample hop. The transform has 257 non-negative frequency bins; we retain bins $0$--$255$ and discard only the Nyquist bin, hence $F{=}256$. At 16\,kHz, the bin width is 31.25\,Hz, so 1, 2, 4, and 8\,kHz fall exactly at bins 32, 64, 128, and 256.

\begin{figure*}[t]
\centering
\includegraphics[width=0.98\textwidth]{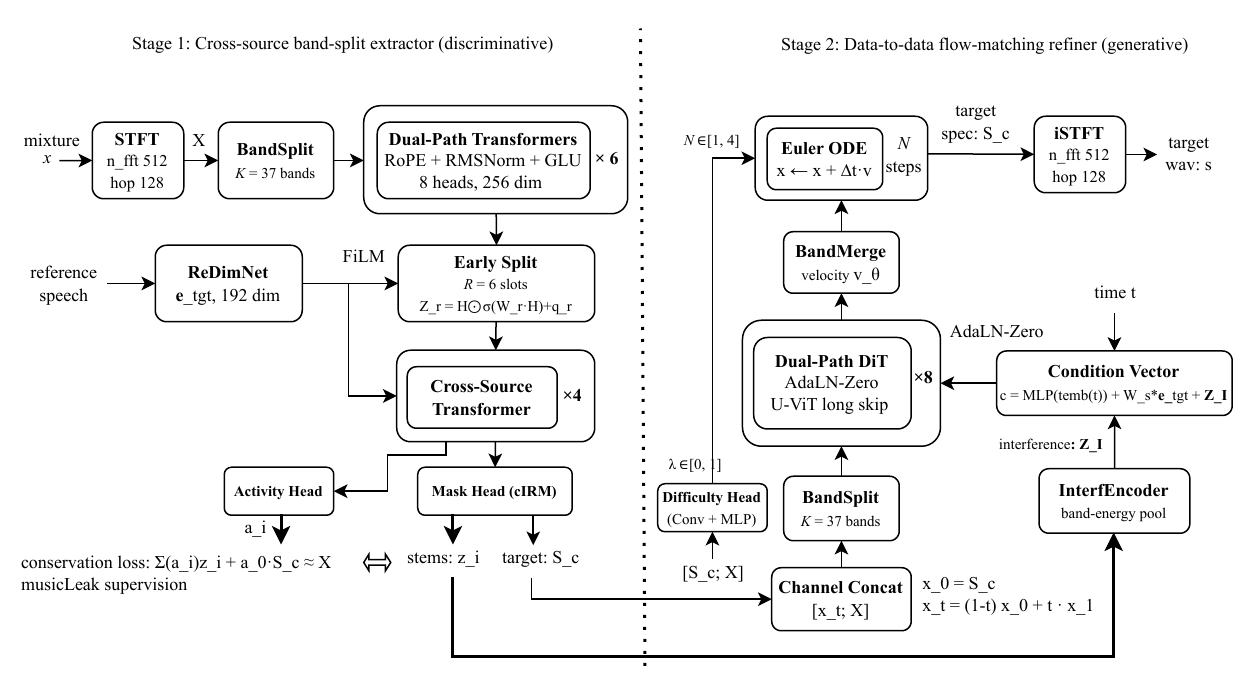}
\caption{MuseFlow-TSE architecture. All spectra use the actual 512-point STFT with bins $0$--$255$ retained ($F{=}256$). \textbf{Stage-1}: band-split ($K{=}37$), $N_b{=}6$ dual-path blocks with RoPE, FiLM-gated early split into $R{=}6$ slots, $N_c{=}4$ cross-source blocks, per-slot cIRMs. \textbf{Stage-2}: 8 AdaLN-Zero DiT dual-path blocks with U-ViT skips over $[x_t;\,X]$, conditioned on the flow time $t$, $\Etgt$, and the negative interference embedding $\zI$; the difficulty head sets the step count $N$ (dashed arrows: conditioning).}
\label{fig:pipeline}
\end{figure*}

A frozen ReDimNet encoder extracts $\Etgt\in\real{192}$ from $e$ and conditions both stages (Fig.~\ref{fig:pipeline}): Stage-1 (discriminative) outputs a coarse target $\hat{S}_{\mathrm{c}}$ plus interference stems $\{\hat{I}_{\mathrm{music}},\hat{I}_{\mathrm{spk}},\hat{I}_{\mathrm{noise}}\}$; a difficulty head $g_\phi$ sets the step count $N$; Stage-2 (generative) performs $N$-step Euler refinement from $\hat{S}_{\mathrm{c}}$, trained with an end-to-end signal-fidelity objective. We compare three configurations: \textbf{Config-D} (Stage-1 only), \textbf{Config-DG} (Stage-1$\to$Stage-2 with adaptive steps), and \textbf{Config-DG-MF} (the MeanFlow one-step variant).

\subsection{Stage-1: Music-Aware Cross-Source Band-Split Extractor}
\textbf{Band-split backbone.} Following BS-RoFormer~\cite{bsroformer}, the $F{=}256$ retained bins are partitioned into $K{=}37$ contiguous non-uniform sub-bands, with nominal resolutions of 2, 4, 12, and 24 bins below 1\,kHz, at 1--2\,kHz, at 2--4\,kHz, and above 4\,kHz, respectively. Let the ordered band-edge set be
\begin{equation}
\begin{aligned}
\mathcal{B}=\{&0,2,4,\ldots,32,\;36,40,\ldots,64,\\
&76,88,100,112,120,128,\;152,176,200,224,240,248,256\}.
\end{aligned}
\label{eq:bands}
\end{equation}
The 37 bands are $[b_k,b_{k+1})$ for consecutive $b_k,b_{k+1}\in\mathcal{B}$; equivalently, the bin-width sequence is $(2^{\times16},4^{\times8},12^{\times4},8^{\times2},24^{\times4},16,8,8)$. This exact tiling uses all retained bins and places the 1/2/4\,kHz boundaries on integer bins. Fine resolution at low frequencies matches the concentration of musical harmonics. Each band is projected to $D{=}256$-d tokens and processed by $N_b{=}6$ dual-path blocks (time then band Transformer; 8 heads, RMSNorm, GLU FFNs, RoPE~\cite{rope,bsroformer}).

\noindent\textbf{Early split and cross-source interaction.} On the backbone output $\mathbf{H}$, $R{=}6$ source slots (1 target $+$ 3 speakers $+$ 1 music $+$ 1 noise) are instantiated by FiLM-gated modulation plus an additive query,
\begin{equation}
\mathbf{Z}_r = \mathbf{H}\odot\sigma(W_r\mathbf{H}) + \mathbf{q}_r,\qquad r=1,\dots,R,
\label{eq:earlysplit}
\end{equation}
The target query is enrollment-conditioned, $\mathbf{q}_{\mathrm{tgt}}=\mathbf{q}_0+\mathrm{MLP}(\Etgt)$; other queries are learnable constants, and singing is folded into the music reference (Sect.~\ref{sec:data}). $N_c{=}4$ three-path blocks apply time, band, and \textbf{cross-source} Transformers, extending SepReformer's cross-speaker attention~\cite{sepreformer} to heterogeneous sources. Identity acts only on the target pathway through its query, an extra cross-attention key/value pair~\cite{speakerbeam}, a tied enrollment/mixture encoder~\cite{spex}, and sparing global AdaLN~\cite{ddtse}, keeping music/noise slots identity-agnostic.

\noindent\textbf{Masks and conservation.} Each slot is decoded by a per-band mask head into a complex mask applied to the mixture, $\hat{S}_r = M_r \odot X$; the target slot yields $\hat{S}_{\mathrm{c}}=\hat{S}_{\mathrm{tgt}}$ and the rest the interference stems. A conservation constraint $\|x-\sum_r\hat{s}_r\|_1$ enforces joint separation, and an activity head supervises slot occupancy under a variable number of speakers.

\subsection{Stage-2: Data-to-Data Flow-Matching Refiner}
\textbf{Data-to-data flow.} Stage-2 \emph{inserts} the detail that masking cannot recover. Following the D/I evidence~\cite{mask2flow}, the flow is anchored at the discriminative estimate: $x_0=\hat{S}_{\mathrm{c}}$, $x_1=S$, $x_t=(1-t)x_0+tx_1$, trained with $\mathcal{L}_{\mathrm{fm}}=\mathbb{E}\|v_\theta(x_t,t\mid X,\Etgt,\zI)-(x_1-x_0)\|^2$~\cite{lipman}. With the start close to the target, few steps suffice and the hallucination space is drastically reduced.

\noindent\textbf{Backbone and conditioning.} The 4-channel $[x_t;\,X]$ is band-split as in Stage-1 and processed by 8 AdaLN-Zero DiT dual-path blocks ($D'{=}384$, 8 heads, GLU FFN $\times4$, RoPE) with U-ViT skips~\cite{dit,soloaudio}. Conditioning combines the mixture reference~\cite{soloaudio,solospeech}, AdaLN-Zero modulation by $\mathbf{c}=\mathrm{MLP}_t(\mathrm{temb}(t))+W_s\Etgt+\zI$~\cite{mask2flow,dit}, and a new negative condition $\zI=W_o[W_m\bar{A}_{\mathrm{music}};W_s'\bar{A}_{\mathrm{spk}};W_n\bar{A}_{\mathrm{noise}}]\in\real{D'}$, obtained by time-averaging and projecting the Stage-1 interference magnitudes, which identifies deleted music/singing that must not return.

\noindent\textbf{Adaptive steps, signal fidelity, one-step variant.} A lightweight difficulty head $g_\phi$ (strided Conv1d, pooling, MLP) estimates the \emph{target energy ratio}
\begin{equation}
\rho=\frac{\|s\|_2^2}{\|s\|_2^2+\sum_j\|u_j\|_2^2+\|m\|_2^2+\|n\|_2^2}\in[0,1]
\label{eq:target-ratio}
\end{equation}
from $(X,\Etgt)$; a larger $\rho$ means an easier, target-dominated mixture. The prediction $\hat{\rho}$ is mapped to $N=\mathrm{clamp}(\mathrm{round}(N_{\min}+(N_{\max}-N_{\min})(1-\hat{\rho})),1,4)$~\cite{adflowtse}: easy mixtures use 1 step, while heavily overlapped musical mixtures use up to 4. An end-to-end $-\mathrm{SI\text{-}SNR}$ term on the gradient-carrying refined output supervises the intrusive metric directly, so no post-corrector is required, and 30\% reference masking of the coarse input~\cite{solospeech} forces the refiner to recover content from the mixture. A parallel MeanFlow variant with $\alpha$-Flow~\cite{meanflow,alphaflow} enables true one-step inference, trained with a larger identity weight~\cite{meanflowtse}. Except for the frozen encoder, all modules are trained from scratch.

\section{Training}
\subsection{Loss Functions}
Let $\mathcal{Q}=\{\mathrm{tgt},1,2,3\}$ denote the four speaker slots (one anchored target and three interfering-speaker slots), with $m$/$n$ the music/singing and noise slots; hence there are six physical slots. For slot $r$, $\hat{s}_r=\mathrm{iSTFT}(\hat{S}_r)$ and $s_r$ is its reference. Stage-1 minimizes
\begin{equation}
\begin{aligned}
\mathcal{L}_1={}&\mathcal{L}_{\mathrm{rec}}
+\alpha(\mathcal{L}_{\mathrm{music}}+\mathcal{L}_{\mathrm{noise}})
+\beta\mathcal{L}_{\mathrm{cons}}
+\lambda_a\mathcal{L}_{\mathrm{mag}}\\
&+\lambda_b\mathcal{L}_{\mathrm{mrstft}}
+\lambda_d\mathcal{L}_{\mathrm{leak}}
+\lambda_{s1}\mathcal{L}_{\mathrm{sid1}}
+\lambda_{n1}\mathcal{L}_{\mathrm{neg1}}
+0.1\mathcal{L}_{\mathrm{act}} .
\end{aligned}
\label{eq:l1}
\end{equation}
$\mathcal{L}_{\mathrm{rec}}$ is grouped-PIT negative SI-SNR over $\mathcal{Q}$ (target anchored, three interfering speakers permuted); $\mathcal{L}_{\mathrm{music}}$/$\mathcal{L}_{\mathrm{noise}}$ are negative SI-SNR to $m$/$n$; $\mathcal{L}_{\mathrm{mag}}$ is normalized magnitude $\ell_1$ error; $\mathcal{L}_{\mathrm{mrstft}}$ uses multi-resolution spectral convergence and log-magnitude errors~\cite{bsroformer}. Conservation and activity are $\mathcal{L}_{\mathrm{cons}}=\|x-\hat{s}_{\mathrm{tgt}}-\sum_{j=1}^{3}\hat{s}_j-\hat{s}_m-\hat{s}_n\|_1$ and binary cross-entropy on active-slot labels, respectively. MusicLeak is $\mathcal{L}_{\mathrm{leak}}=|\alpha_m|^2\|m\|_2^2/(\|\hat{s}_{\mathrm{tgt}}\|_2^2+\epsilon)$, where $\alpha_m=\langle\hat{s}_{\mathrm{tgt}},m\rangle/(\|m\|_2^2+\epsilon)$. With frozen ReDimNet $E$, $\mathcal{L}_{\mathrm{sid1}}=1-\cos(E(\hat{s}_{\mathrm{tgt}}),\Etgt)$ pulls toward enrollment and $\mathcal{L}_{\mathrm{neg1}}=\sum_{e^-}\max(0,\cos(E(\hat{s}_{\mathrm{tgt}}),e^-))$ pushes away from interfering speech and singing hard negatives.

Stage-2 minimizes
\begin{equation}
\mathcal{L}_2=\mathcal{L}_{\mathrm{fm}}
+\lambda_{\mathrm{sid}}\mathcal{L}_{\mathrm{sid}}
+\lambda_{\mathrm{neg}}\mathcal{L}_{\mathrm{neg}}
+\lambda_{\mathrm{sig}}\mathcal{L}_{\mathrm{sig}}
+\lambda_{\mathrm{mr}}\mathcal{L}_{\mathrm{mrstft}}
+\mathcal{L}_g .
\label{eq:l2}
\end{equation}
The identity terms mirror Stage-1, $\mathcal{L}_{\mathrm{sig}}=-\mathrm{SI\text{-}SNR}$, and both, together with $\mathcal{L}_{\mathrm{mrstft}}$, are evaluated on the gradient-carrying few-step Euler output rather than $\hat{x}_1$. The difficulty-head loss is $\mathcal{L}_g=(\hat{\rho}-\rho)^2$. All $\lambda$'s above are scalar loss weights, distinct from target ratio $\rho$. Initial weights: $\alpha{=}1.0$, $\beta{=}\lambda_a{=}\lambda_d{=}0.5$, $\lambda_b{=}1.0$, $\lambda_{s1}{=}\lambda_{n1}{=}0.3$ (Stage-1); $\lambda_{\mathrm{sid}}{=}1.0$ (larger for MeanFlow), $\lambda_{\mathrm{neg}}{=}\lambda_{\mathrm{mr}}{=}0.5$, and $\lambda_{\mathrm{sig}}$ ramps from $0.2$ to $0.3$ (Stage-2).

\subsection{Data Recipe}
\label{sec:data}
Three-second mixtures are synthesized online from LibriSpeech, VoxCeleb2, and WSJ0 speech (2--4 total speakers, SNR $-5\sim5$\,dB)~\cite{librispeech,voxceleb2}, MUSDB18-HQ and MoisesDB music (prob.\ up to 0.7, SNR $-10\sim10$\,dB)~\cite{musdb18,moisesdb}, and noise (prob.\ up to 0.9, SNR $0\sim15$\,dB)~\cite{wham}, with optional reverberation. Singing serves as music-slot ground truth and as a hard-negative embedding; enrollments are augmented with probability 0.7. All $R{=}6$ physical slots remain fixed; the curriculum changes only active speaker slots and interference probabilities: (i)~two active speakers, light noise (0.5), no music/singing; (ii)~up to three speakers, music/singing (0.5/0.3), noise (0.8); (iii)~up to four speakers, music/singing/noise (0.7/0.5/0.9). Inactive slots are supervised toward silence.

\subsection{Staged Training Strategy}
Training uses four sequential stages with the speaker encoder frozen: T1 trains Stage-1 through the curriculum, initializing each level from the previous one; T2 freezes Stage-1 and trains the flow, using actual Stage-1 predictions as starts with probability $\min(0.45,\mathrm{epoch}/100)$ to close the train/inference gap~\cite{ddtse}; T2$'$ trains MeanFlow; and optional T3 fine-tunes only cross-source blocks, output heads, and Stage-2. All use AdamW (lr $2{\times}10^{-4}$, cosine decay, bf16, EMA 0.999).

\section{Experiments}
\subsection{Experimental Setup}
\label{sec:setup}
\textbf{Setup.} The 16\,kHz model uses Sect.~3's configuration. Experiments use the Complex scenario---online clean/noisy/music/singing/reverberant mixtures with disjoint lists. This controlled synthetic benchmark supports design attribution but does not establish natural-recording generalization; stratified and real-world tests remain necessary.

\noindent\textbf{Metrics}: SI-SDR/SI-SDRi~\cite{sisdr}, PESQ~\cite{pesq}, DNSMOS~\cite{dnsmos}, SIM, WER (Whisper~\cite{whisper}), MusicLeak, singing leakage, and NFE. Baselines are SpEx+~\cite{spex}, USEF-TSE~\cite{useftse}, adapted TF-GridNet and SepReformer~\cite{tfgridnet,sepreformer}, DPM-TSE~\cite{dpmtse}, SoloSpeech~\cite{solospeech}, AD-FlowTSE~\cite{adflowtse}, MeanFlow-TSE~\cite{meanflowtse}, Mask2Flow-TSE~\cite{mask2flow}, and a BS-RoFormer$\to$SepReformer cascade.

\subsection{Comparison Experiments}
Table~\ref{tab:main} compares all systems. Config-D attains the best WER (12.0\,\%) and MusicLeak (2.9\,\%) with SIM 0.89, confirming content faithfulness. Config-DG then improves SI-SDRi to \textbf{13.2}\,dB and PESQ to \textbf{2.88}, with SIM rising to \textbf{0.90} and WER essentially preserved (12.2\,\%), indicating refinement without identity drift. TF-GridNet leaks 2.5$\times$ more music energy than Config-D (7.2\,\% vs.\ 2.9\,\%), while the BS-RoFormer$\to$SepReformer cascade trails Config-D by 2.5\,dB SI-SDRi with 7.8\,\% MusicLeak, showing that direct cascading cannot replace constrained joint estimation.

\begin{table}[t]
\caption{Main comparison (Complex scenario; WER/MusicLeak in \%). SoloSpeech: 50 steps, 8 candidates.}
\label{tab:main}
\centering
\scriptsize
\setlength{\tabcolsep}{3pt}
\begin{tabular}{@{}lccccccc@{}}
\toprule
Method & SI-SDRi$\uparrow$ & PESQ$\uparrow$ & DNSMOS$\uparrow$ & SIM$\uparrow$ & WER$\downarrow$ & MusicLeak$\downarrow$ & NFE$\downarrow$ \\
\midrule
SpEx+~\cite{spex} & 8.2 & 2.02 & 2.55 & 0.76 & 22.8 & 9.7 & 1 \\
USEF-TSE~\cite{useftse} & 9.1 & 2.18 & 2.68 & 0.79 & 20.5 & 8.9 & 1 \\
TF-GridNet~\cite{tfgridnet} & 11.9 & 2.58 & 2.95 & 0.85 & 14.6 & 7.2 & 1 \\
DPM-TSE~\cite{dpmtse} & 9.6 & 2.35 & 3.05 & 0.77 & 19.8 & 8.4 & 50 \\
SoloSpeech~\cite{solospeech} & 10.4 & 2.72 & \textbf{3.28} & 0.84 & 15.9 & 6.8 & 50$\times$8 \\
AD-FlowTSE~\cite{adflowtse} & 10.8 & 2.66 & 3.18 & 0.83 & 16.4 & 7.0 & ${\sim}$8 \\
MeanFlow-TSE~\cite{meanflowtse} & 10.2 & 2.61 & 3.12 & 0.71 & 18.2 & 7.9 & 1 \\
Mask2Flow-TSE~\cite{mask2flow} & 11.5 & 2.74 & 3.21 & 0.84 & 15.2 & 6.5 & 1 \\
Direct Cascade~\cite{bsroformer,sepreformer} & 10.5 & 2.40 & 2.80 & 0.81 & 17.5 & 7.8 & 1 \\
\midrule
Config-D (ours) & 13.0 & 2.71 & 3.08 & 0.89 & \textbf{12.0} & \textbf{2.9} & 1 \\
Config-DG (ours) & \textbf{13.2} & \textbf{2.88} & 3.06 & \textbf{0.90} & 12.2 & 3.4 & 1.6 \\
Config-DG-MF (ours) & 12.6 & 2.79 & 3.01 & 0.86 & 13.4 & 4.6 & 1 \\
\bottomrule
\end{tabular}
\end{table}

\paragraph{When is the refinement stage useful?}
D$\to$DG is a quality--latency trade-off, not a universally necessary second stage: SI-SDRi rises by 0.2\,dB, while PESQ improves more noticeably (2.71$\to$2.88), SIM/WER are preserved, and average NFE rises from 1 to 1.6. Thus Config-D suits latency-critical recognition, Config-DG suits perceptual reconstruction, and Config-DG-MF is the lower-quality one-step compromise.

\subsection{Ablation Designs and Component Attribution}
Every removal in Table~\ref{tab:ablation} degrades its primary metric, with $\zI$ and the cross-source axis most important. Music-specific attribution is direct: removing $\zI$ raises MusicLeak/singing leakage from 3.4\%/2.1\% to 6.9\%/5.4\%; removing band-splitting raises MusicLeak from 2.9\% to 6.2\% and reduces music SDR from 9.4 to 6.1\,dB. These component-level results support the music-aware design, but the aggregated Complex scenario still does not replace stratified evaluation. The D/I re-analysis confirms the division of labor, and adaptive steps average 1.6 NFE at equal DNSMOS.

\begin{table}[t]
\caption{Ablation results on the Complex scenario (primary metrics; values are before$\to$after).}
\label{tab:ablation}
\centering
\scriptsize
\begin{tabularx}{\textwidth}{@{}cl>{\hsize=.6\hsize\raggedright\arraybackslash}X>{\hsize=1.4\hsize\raggedright\arraybackslash}X@{}}
\toprule
\# & Modification & Validates & Observed \\
\midrule
1 & Config-D $\to$ DG & drift-free refinement & SI-SDRi $+0.2$\,dB; PESQ $+0.17$; SIM $+0.01$; DNSMOS preserved \\
2 & Gaussian $\to$ $\hat{S}_{\mathrm{c}}$ start & start point & NFE 16$\to$1.6; WER 16.8$\to$12.2\,\% \\
3 & w/o cross-source TF & decoupling & SI-SDRi 13.0$\to$11.2\,dB; cons.\ err.\ 0.021$\to$0.048 \\
4 & w/o negative condition $\zI$ & anti-hallucination & MusicLeak 3.4$\to$6.9\,\%; singing 2.1$\to$5.4\,\% \\
5 & w/o band-split & music modeling & MusicLeak 2.9$\to$6.2\,\%; music SDR 9.4$\to$6.1\,dB \\
6 & w/o conservation / grouped PIT & joint estimation & SI-SDRi $-1.2$\,dB / $-0.9$\,dB \\
7 & fixed $\to$ adaptive $N$ & adaptive compute & NFE 4.0$\to$1.6 at equal DNSMOS \\
8 & w/o $\mathcal{L}_{\mathrm{sid}}$ / $\mathcal{L}_{\mathrm{sig}}$ & anti-drift & SIM 0.90$\to$0.79; SI-SDRi $-0.9$\,dB \\
9 & D/I re-analysis & D/I split & 91:9 (Stage-1), 38:62 (Stage-2 step-1) \\
\bottomrule
\end{tabularx}
\end{table}

\section{Conclusion}
We presented MuseFlow-TSE, combining a music-aware cross-source band-split extractor with a negatively conditioned, difficulty-adaptive data-to-data flow refiner. Evidence remains limited to 16\,kHz synthetic mixtures: the music-specific ablations support the mechanisms but do not establish natural-recording generalization. Future work will report stratified scenario results, real-world benchmarks, and perception-aligned fine-tuning.

\end{document}